%% file: main.tex
\documentclass[sigconf,10pt]{acmart}

\usepackage{graphicx}
\usepackage{caption}
\usepackage{subfigure}

\usepackage{listings}
\usepackage{enumitem}
\setlist[itemize]{align=parleft,left=0pt..1em}

\usepackage{acronym}

\usepackage{comment}
\usepackage{wrapfig}
\usepackage{tikz}
\usepackage{xcolor}

\setcopyright{acmcopyright}

\acmDOI{10.475/1234.5678}

\acmISBN{123-4567-24-567/08/06}

\acmConference[HotStorage'21]{The 13th ACM Workshop on Hot Topics in Storage and File Systems}{July 27-28, 2021}{Virtual}
\acmYear{2021}
\copyrightyear{2021}

\acmPrice{15.00}

\begin{document}
\title{Towards an Arrow-native Storage System}

\author{Jayjeet Chakraborty}
\affiliation{%
  \institution{UC Santa Cruz}
  \city{Santa Cruz}
  \state{California, USA}
}
\email{jchakra1@ucsc.edu}

\author{Ivo Jimenez}
\affiliation{%
  \institution{UC Santa Cruz}
  \city{Santa Cruz}
  \state{California, USA}
}
\email{ivotron@ucsc.edu}

\author{Sebastiaan Alvarez Rodriguez}
\affiliation{%
  \institution{Leiden University}
  \city{Leiden}
  \state{Netherlands}
}
\email{s.f.alvarez.rodriguez@umail.leidenuniv.nl}

\author{Alexandru Uta}
\affiliation{%
  \institution{Leiden University}
  \city{Leiden}
  \state{Netherlands}
}
\email{a.uta@liacs.leidenuniv.nl}

\author{Jeff LeFevre}
\affiliation{%
  \institution{UC Santa Cruz}
  \city{Santa Cruz}
  \state{California, USA}
}
\email{jlefevre@ucsc.edu}

\author{Carlos Maltzahn}
\affiliation{%
  \institution{UC Santa Cruz}
  \city{Santa Cruz}
  \state{California, USA}
}
\email{carlosm@ucsc.edu}

\renewcommand{\shortauthors}{J. Chakraborty et al.}

\begin{abstract}
\input{tex/1_abstract}
\end{abstract}


%
%

\maketitle

\sloppy

\input{tex/2_intro}
\input{tex/3_design}
\input{tex/4_evaluation}
\input{tex/5_related}
\input{tex/6_conclusion}

\bibliographystyle{bib/ACM-Reference-Format}
\bibliography{bib/main}

\end{document}

%% file: tex/1_abstract.tex
With the ever-increasing dataset sizes, several file formats like Parquet, ORC, and Avro have been developed to store data efficiently and to save network and interconnect bandwidth at the price of additional CPU utilization. However, with the advent of networks supporting $25$-$100$ Gb/s and storage devices delivering $1,000,000$ reqs/sec the CPU has become the bottleneck, trying to keep up feeding data in and out of these fast devices. The result is that data access libraries executed on single clients are often CPU-bound and cannot utilize the scale-out benefits of distributed storage systems. One attractive solution to this problem is to offload data-reducing processing and filtering tasks to the storage layer. However, modifying legacy storage systems to support compute offloading is often tedious and requires extensive understanding of the internals. Previous approaches re-implemented functionality of data processing frameworks and access library for a particular storage system, a duplication of effort that might have to be repeated for different storage systems.

In this paper, we introduce a new design paradigm that allows extending programmable object storage systems to embed existing, widely used data processing frameworks and access libraries into the storage layer with minimal modifications. In this approach data processing frameworks and access libraries can evolve independently from storage systems while leveraging the scale-out and availability properties of distributed storage systems. We present one example implementation of our design paradigm using Ceph, Apache Arrow, and Parquet. We provide a brief performance evaluation of our implementation and discuss key results.

\begin{figure}[h]
\centering
\includegraphics[width=0.9\linewidth]{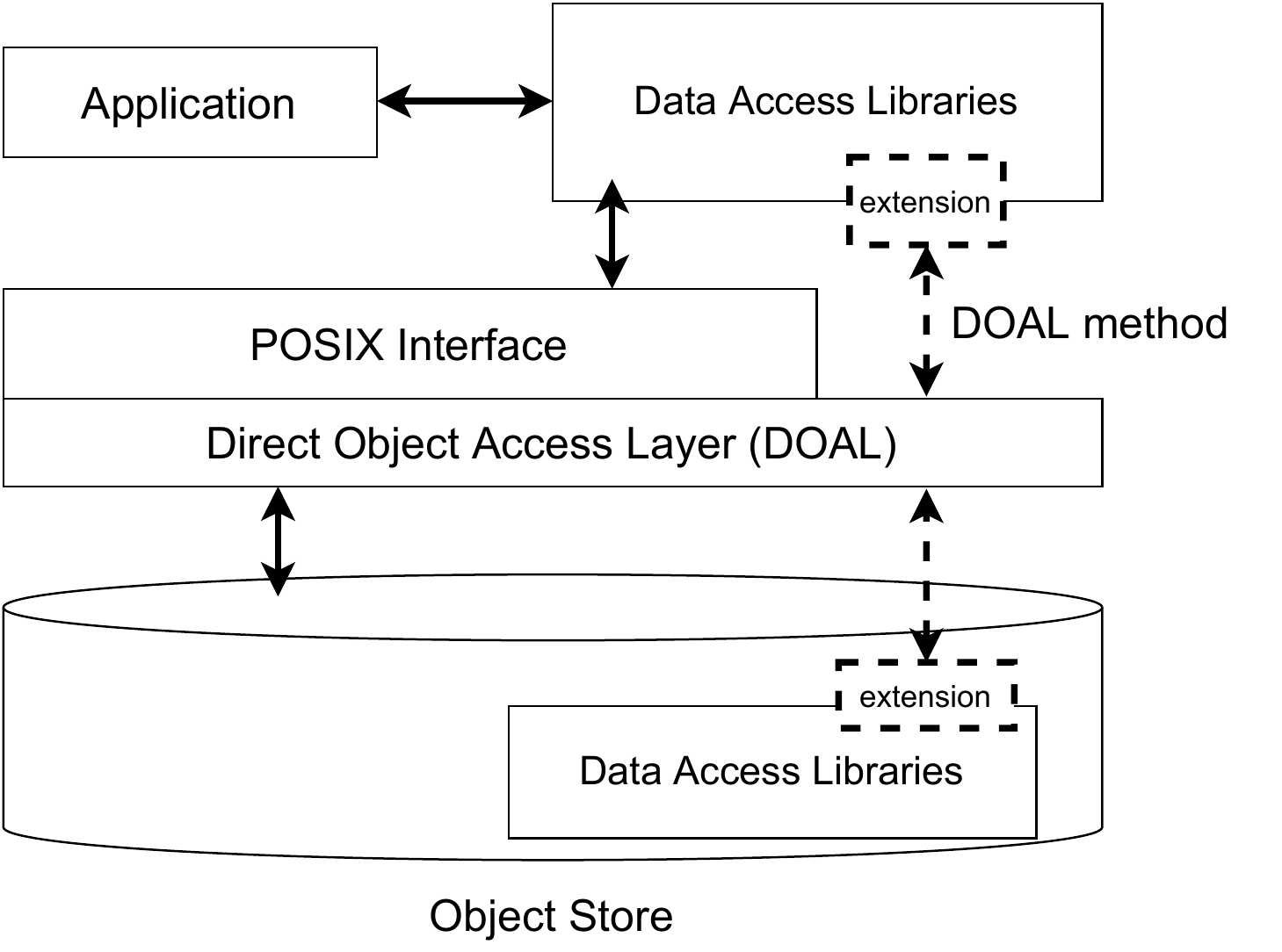}
\caption{High-Level Design and Architecture.}
\label{fig:hld}
\end{figure}

%% file: tex/2_intro.tex
\section{Introduction}

Over the last decade, a variety of distributed data processing frameworks like Spark~\cite{zaharia2010spark} and Hadoop~\cite{white2012hadoop} have come into existence. These frameworks were built to efficiently query vast quantities of semi-structured data and get insights quickly. Unlike standard relational database management systems (RDBMS) such as MySQL~\cite{mysql}, which are optimized to manage both data storage and processing, these systems were designed to read data from a wide variety of data sources, including those in the cloud like S3~\cite{S3}. These systems depend on different file formats like Parquet~\cite{Parquet}, Avro~\cite{Avro}, and ORC~\cite{ORC} for efficiently storing and accessing data. Since storage devices have been the primary bottleneck for data processing systems for a long time, the main focus of these file formats has been to store data efficiently in a binary format and reduce the amount of disk I/O required to fetch the data. However, with recent advancements in storage devices with NVMe~\cite{xu2015performance} drives and network devices with Infiniband networks~\cite{infiniband}, the bottleneck has shifted from the storage devices to the client machine's CPUs, rendering the notion of "A fast CPU and slow disk" invalid as shown by Trivedi et al. ~\cite{trivedi2018albis}. The serialized and compressed nature of these file formats makes reading them CPU bound in systems with high-speed network and storage devices, resulting in severely reduced scalability.

An attractive solution to this problem is to offload any computation to the storage layer to achieve scalability, faster queries, and less network traffic. Several popular distributed data processing systems have explored this approach, e.g. IBM Netazza~\cite{singh2011introduction}, Oracle Exadata~\cite{OracleExadata}, Redshift~\cite{gupta2015amazon}, and PolarDB~\cite{cao2020polardb}. Most of these systems are built following a clean-slate approach and use specialized and costly hardware, such as Smart SSDs~\cite{do2013query} and FPGAs~\cite{fpga} for table scanning. Building systems like these requires in-depth understanding and expertise in building database systems. Also, modifying existing systems like MariaDB~\cite{razzoli2014mastering}, as in the case of YourSQL~\cite{jo2016yoursql}, requires modifying code that is hardened over the years which may result in performance, security, and reliability issues. A possible way to mitigate these issues is to have programmable storage systems with low-level extension mechanisms that allow implementing application-specific data manipulation and access in their I/O path. Customizing storage systems via plugins results in minimal implementation overhead and increases the maintainability of the software. 

Programmable object-storage systems like Ceph~\cite{weil2006ceph}, Swift~\cite{swift}, and DAOS~\cite{liang2020daos} often provide a POSIX filesystem interface for reading and writing files which are mostly built on top of direct object access libraries like ``librados'' in Ceph and ``libdaos'' in DAOS. Being programmable, these systems provide plugin-based extension mechanisms that allow direct access and manipulation of objects within the storage layer. We leverage these features of programmable storage systems and develop a new design paradigm that allows the embedding of widely-used data access libraries inside the storage layer. As shown in Figure~\ref{fig:hld}, the extensions on the client and storage layers allow an application to execute access library operations either on the client or via the direct object access layer, in the storage server.

We implement one instantiation of our design paradigm using RADOS~\cite{weil2007rados} as the object-storage backend, CephFS~\cite{borges2017cephfs} as the POSIX layer, Apache Arrow~\cite{Arrow,ArrowDatasetDocs} as the data access library, and Parquet as the file format. We evaluate the performance of our implementation by varying parameters such as cluster size and parallelism and measure metrics like query latency and CPU utilization. The evaluations show that our implementation scales almost linearly by offloading CPU usage for common data processing tasks to the storage layer, freeing the client for other processing tasks. 

In summary, our primary contributions are as follows:
\begin{itemize}[leftmargin=*]
\item A design paradigm that allows extending programmable storage systems with the ability to offload CPU-bound tasks like dataset scanning to the storage layer using plugin-based extension mechanisms and widely-used data access libraries while keeping the implementation overhead minimal.

\item A brief analysis of the performance gained by offloading Parquet dataset scanning to the storage nodes. We demonstrate that offloading dataset scanning operations to the storage layer results in faster queries and near-linear scalability.
\end{itemize}

%% file: tex/3_design.tex
\section{Design and Implementation}
\label{sec:design_and_impl}
In this section, we discuss our design paradigm, the motivation behind it, and provide an in-depth discussion of the internals of our implementation. Additionally, we discuss two alternate file layout choices for efficiently storing and querying Parquet files in Ceph.

\begin{figure*}[h]
\centering
\includegraphics[width=0.9\textwidth]{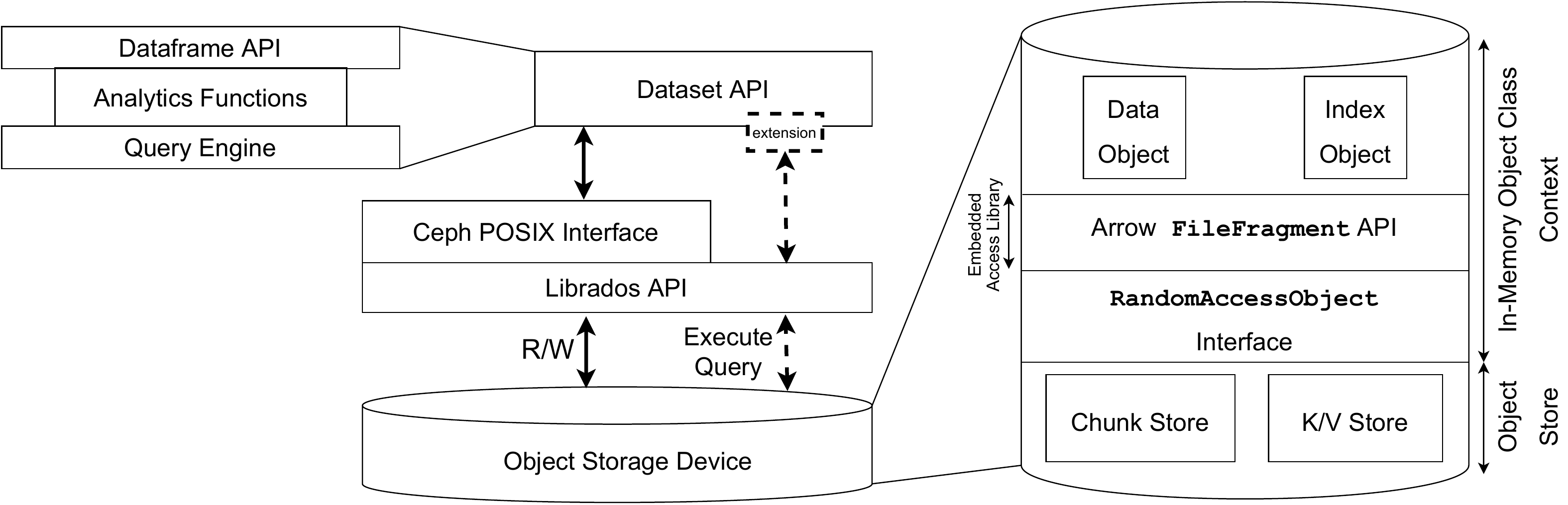}
\caption{Architecture of the implementation that extends CephFS to allow invoking methods for processing Parquet files inside Ceph OSDs using Arrow libraries.}
\label{fig:architecture}
\end{figure*}

\subsection{Design Paradigm}
\label{sec:design_paradigm}
One of the most important aspects of our design is that it allows building in-storage data processing systems with minimal implementation effort. Our design allows extending the client and storage layers with widely-used data access libraries requiring minimal modifications. We achieve this by (1) creating a file system shim in the object storage layer so that access libraries embedded in the storage layer can continue to operate on files, (2) mapping client requests-to-be-offloaded directly to objects using file system striping metadata, and (3) mapping files to logically self-contained fragments by using standard file system striping. As shown in Figure~\ref{fig:architecture}, We developed one instantiation of our design paradigm using Ceph as the storage system, Apache Arrow as the data access library, and Parquet as the file format. We expose our implementation via the Arrow Dataset API by creating a new file format called RADOS Parquet that inherits from the Parquet file format in Arrow. Next, we discuss the internals of our implementation in detail.

\subsection{An Example Implementation}
\label{sec:implementation}
\textbf{Extending Ceph Object Store.}
Inside the storage layer, using the Ceph Object Class SDK~\cite{objectclasssdk}, we create a ``scan\_op'' Object Class method that reads objects containing Parquet binary data from RADOS, scans them using Apache Arrow, and returns the results in the form of an Arrow table. Reading Parquet files requires a random access interface to seek around in the file and read data at particular offsets. Since the Object Class SDK only provides primitives for doing I/O at a particular offset within an object, we utilize these primitives to create a RandomAccessObject interface that provides a file-like view of an object. Arrow provides a FileFragment API that wraps a file and allows scanning it. It takes predicates and projections as input and applies them on Parquet files to return filtered and projected Arrow tables. Since the RandomAccessObject interface allows interacting with objects as files, it plugs into the FileFragment API seamlessly.

\textbf{Extending Ceph Filesystem.} 
The Ceph filesystem provides a POSIX interface and stripes files over objects stored in the RADOS layer. To execute the ``scan\_op'' Object Class method on a Parquet file, CephFS first needs to map the file to object IDs. CephFS provides access to metadata information on how files in CephFS are mapped to objects in RADOS. We leverage this information to derive object IDs from filenames. We implement a DirectObjectAccess API that facilitates this translation and allows calling Object Class methods like ``scan\_op'' on files. With this API, we obtain the ability to interact with RADOS-stored objects and manipulate them directly in application-specific ways, while also having a filesystem view over the objects.

\textbf{Extending Arrow Dataset API.} Arrow provides a FileFormat API~\cite{ArrowFileFormat} that plugs into the Dataset API~\cite{ArrowDatasetDocs} and allows scanning datasets of different formats in a unified manner. Since Parquet is the de-facto file format of choice in data processing systems, we use it as a baseline for our implementation. We extend the ParquetFileFormat API in Arrow to create a RadosParquetFileFormat API, that allows offloading Parquet file scanning to the Ceph Object Storage Devices (OSDs). This allows client applications using the Dataset API to offload Parquet file scan operations to the Ceph storage layer, by simply changing the file format argument in the Dataset API. 

\subsection{File Layout Designs}
\label{sec:file_layout_design}
Parquet, being very efficient in storing and accessing data, has become the de-facto file format for popular data processing systems like Spark and Hadoop. Since Parquet files are often multiple gigabytes in size, a standard way to store Parquet files is to store them in blocks as in HDFS~\cite{borthakur2007hadoop,HDFS}, where a typical block size is 128MB~\cite{hdfsblocksize}. While writing Parquet files to HDFS, each row group is stored in a single block to prevent reading across multiple blocks when accessing a single row group. We aim to follow a similar file layout for storing Parquet files in Ceph so that every row group is self-contained within an object. In this section, we explore two different approaches for storing Parquet files in Ceph in an HDFS like manner. We call them the Striped and Split file designs. 

\begin{figure}[h]
\centering
\includegraphics[width=\linewidth]{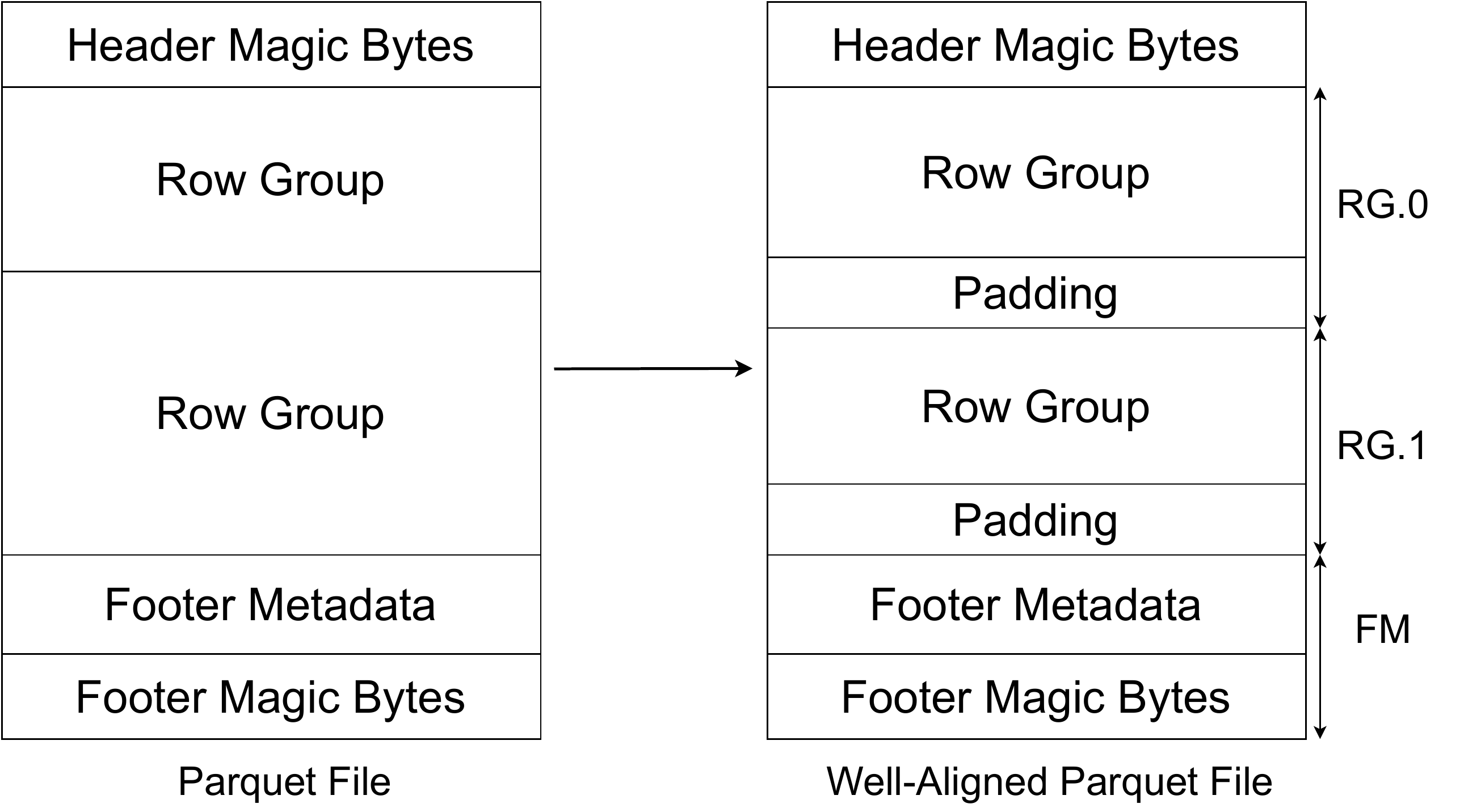}
\caption{\emph{The Striped File Design:} Parquet files are rewritten by padding each row group to create files with equal-sized and object-aligned row groups.}
\label{fig:striped}
\end{figure}

\textbf{Striped File Design.}
In this design, we take a Parquet file and rewrite it by padding its row groups to make them equal-sized. Then on writing the file to CephFS, the RADOS Striper automatically stripes the file into fixed-sized objects. We configure the stripe unit size as a multiple of the row group size, such that a row group is never split across stripe units. Crucially, this allows complete rows and their self-contained semantics to reside within objects. The client maintains a map of the row group and the object ID which contains the row group. During query execution, for every file in CephFS, the filename is first translated into the associated object IDs that constitute the file using the DirectObjectAccess API, and the last object which is expected to contain the Parquet file footer is read. The metadata in a Parquet file footer contains row group statistics that allow the Dataset API to find out the row groups that need to be scanned in the Parquet file based on the query predicates. This capability of Parquet is called `predicate pushdown'. Once the row groups that need to be scanned are calculated, the row group IDs are converted to objects IDs from the map that the client maintains and scan operations are launched in parallel on each of these objects. The resulting Arrow tables from all the scanned objects are materialized into the final result table in the client. The striped file design is shown in Figure~\ref{fig:striped}.

\begin{figure}[h]
\centering
\includegraphics[width=\linewidth]{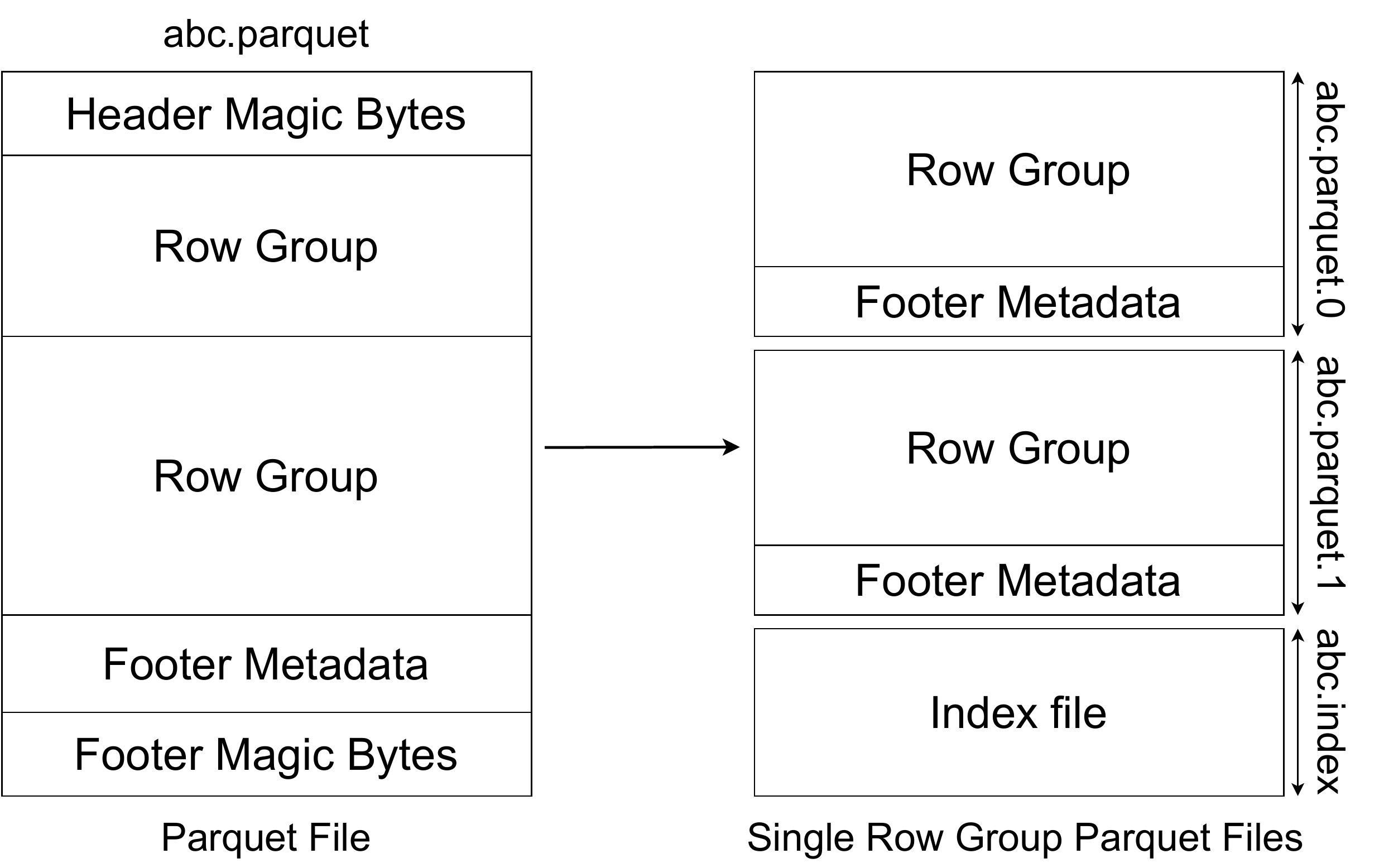}
\caption{\emph{The Split File Design:} Parquet files are split into multiple smaller Parquet files and are accompanied by index files for preserving Parquet optimizations.}
\label{fig:split}
\end{figure}

\textbf{Split File Design.}
In this design, a Parquet file with $R$ row groups is taken and is split into $R$ Parquet files each containing the data from a single row group. To not lose the optimizations due to the predicate pushdown capability in Parquet, the footer metadata and the schema of the parent Parquet file are serialized and written to a separate file ending with a ".index" extension. So, for every Parquet file containing $R$ row groups, we write $R + 1$ small Parquet files each containing a single row group. During the dataset discovery phase, we discover only those files that end with a ".index" extension, translate the filenames to the underlying object IDs, and read the schema information using a RADOS Object Class method. At the start of the query execution phase, the index files are scanned to get the footer metadata. Then, the IDs of the row groups that qualify for scanning are calculated based on the row group statistics present in the footer metadata. Finally, the row group IDs are converted to the corresponding filenames and the underlying objects of these files are then scanned in parallel via the DirectObjectAccess API. Figure~\ref{fig:split} shows the split file design.

%% file: tex/4_evaluation.tex
\section{Evaluation}

We perform experiments to compare the query duration and CPU utilization of accessing a dataset, filtering at the client or at the storage server. Our experiments were performed on CloudLab, the NSF-funded bare-metal-as-a-service infrastructure~\cite{Duplyakin+:ATC19}. For our experiments, we exclusively used machines with an 8-core Intel Xeon D-1548 $2.0$ GHz processor (with hyperthreading enabled), $64$\,GB DRAM, a $256$\,GB NVMe drive, and a $10$\,GbE network interface. These bare-metal nodes are codenamed `m510' in CloudLab. We ran our experiments on Ceph clusters with $4$, $8$, and $16$ storage nodes with a single client. Each storage node had a single Ceph OSD running on top of the NVMe drive. The OSDs were configured to use $8$ threads to prevent any contention due to hyperthreading in the storage nodes. A CephFS was created on a 3-way replicated pool and was mounted in user mode using the ceph-fuse utility.

Our workload comprised of a dataset with $1.2$ billion rows and $17$ columns consisting of data from the NYC yellow taxi dataset~\cite{yellowtaxi}. The in-memory size of the dataset was found to be $154.8$\,GB. In this work, we experiment with $64$\,MB files as we found out that Parquet was more efficient in scanning a few large files than a large number of small files. As described in section~\ref{sec:file_layout_design}, since a row group is supposed to be self-contained within a single object and the unit of parallelism in the Arrow Dataset API when using Parquet is a single row group, we used Parquet files having a single row group backed by a single RADOS object in all our experiments. We use the split file layout design except for the index file with both Parquet and RADOS Parquet. For the experiments, we used the Python version of the Arrow Dataset API and utilized Python's ThreadPoolExecutor for launching scans in parallel, using an Asynchronous I/O model. We measured latency, scalability, and CPU usage for both Parquet and RADOS Parquet.

\begin{figure}[h]
\centering
\includegraphics[width=\linewidth]{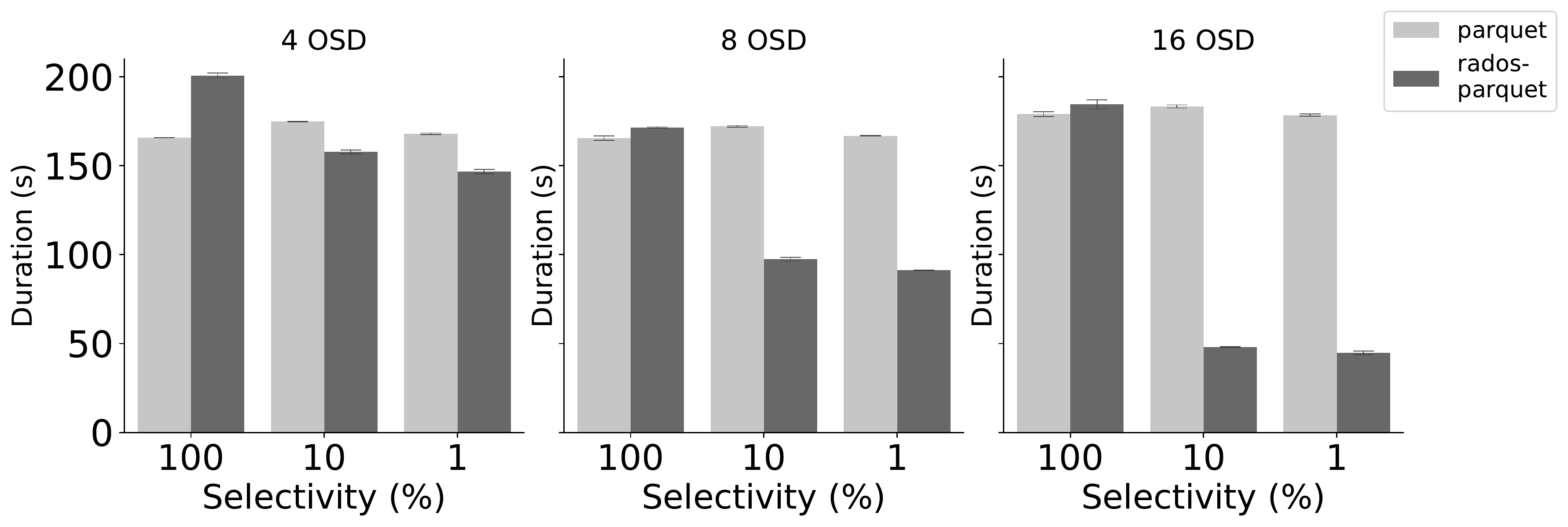}
\caption{Query latency improvement on scaling out from 4 to 8 and 16 storage nodes.}
\label{fig:latency}
\end{figure}

We ran queries to select $100\%$, $10\%$, and $1\%$ of the rows from our dataset. In the $100\%$ selectivity case, all the rows are returned without applying any filters on the dataset. The queue depth at each storage node was maintained at 4 across all the experiments. As shown in Figure~\ref{fig:latency}, RADOS Parquet is faster than Parquet in the $10\%$ and $1\%$ scenarios. Since RADOS Parquet transfers data in the much larger Arrow format as compared to the serialized binary format in the case of Parquet, the $100\%$ selectivity case bottlenecks on the $10$\,GbE network on scaling out, resulting in no performance improvement. Except for the $100\%$ selectivity case, on scaling out from $4$ to $16$ OSDs, RADOS Parquet keeps getting faster than Parquet due to its ability to offload and distribute the computation across all the storage nodes whereas Parquet remains CPU bottlenecked on the client. 

\begin{figure}[h]
\centering
\includegraphics[width=\linewidth]{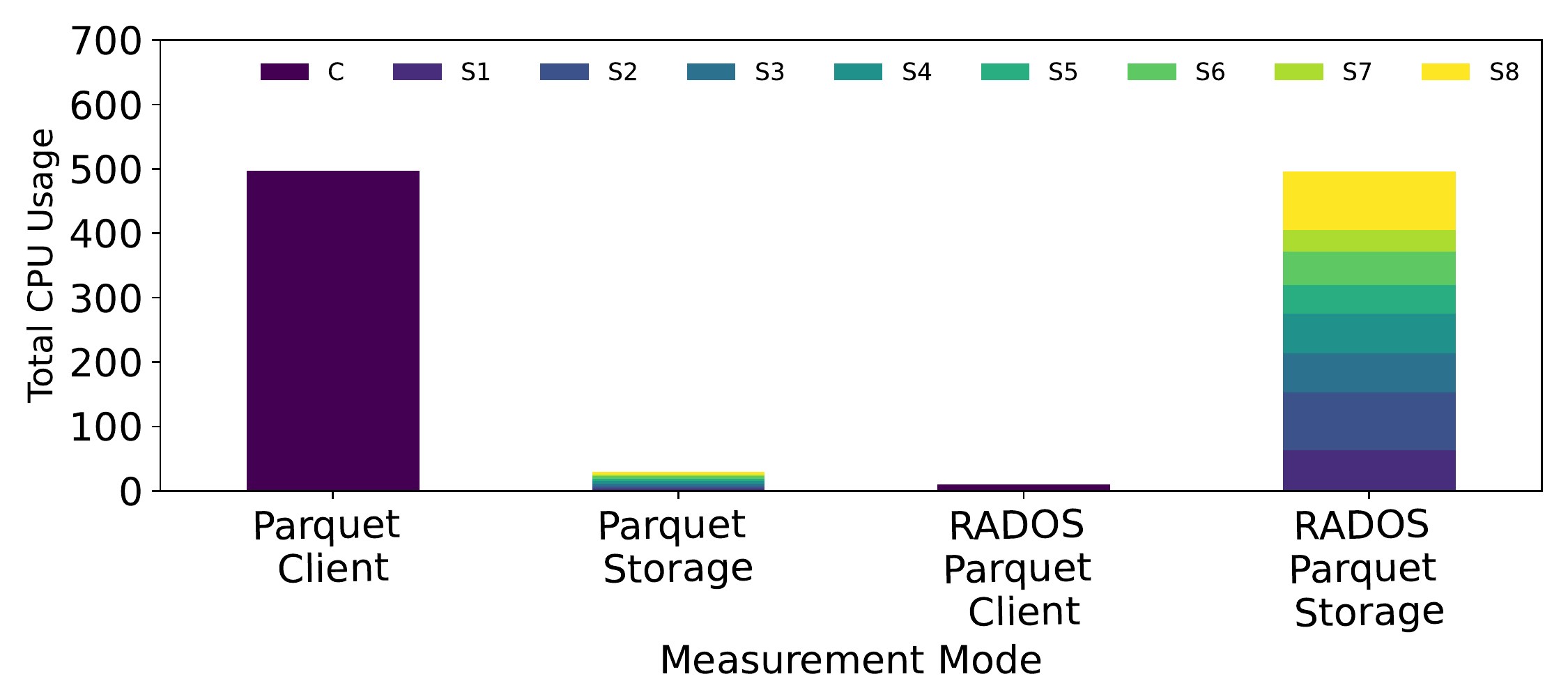}
\caption{Total CPU utilization by Parquet and RADOS Parquet on the client and storage nodes for 60 seconds (measured at 15s intervals) during a query execution with 100\% selectivity. Here C represents the client node while S[1-8] represent the storage nodes.}
\label{ref:cpu}
\end{figure}

Figure~\ref{ref:cpu} shows the total CPU utilization for $60$ seconds by the client and the storage layer during a $100\%$ selectivity query execution with $8$ storage nodes and $16$ client threads. We observe that even with no filtering involved, Parquet exhausts the client's CPU. This implies that the client would be unable to do any other processing work in this situation. On the other hand, we observe that with RADOS Parquet, the client's CPU is almost idle and all the CPU usage is at the storage layer. Hence, with the client's CPU almost free, more asynchronous threads can be launched to improve parallelism or the client can do other processing tasks.

%% file: tex/5_related.tex
\section{Related Work}
Several distributed data processing systems have embraced the idea of query offloading to the storage layer for performance improvement. The paper on the Albis~\cite{trivedi2018albis} file format by Trivedi et al. explored that with high-performance storage and networks, CPUs have become the new bottleneck. Since the CPU bottleneck on the client hampers scalability, offloading CPU to the storage layer has become more important. Recently, S3 introduced S3 Select~\cite{s3select}, which allows files in either Parquet, CSV, or JSON format to be scanned inside S3 for improved query performance. Many other systems which already support reading from S3, e.g. Spark~\cite{Spark}, Redshift~\cite{gupta2015amazon}, Snowflake~\cite{Snowflake}, and PushdownDB~\cite{yu2020pushdowndb} have started taking advantage of S3 Select. But being an IaaS~\cite{iaas}, the performance of S3 Select cannot be tuned, nor can it be customized to read from file formats except the ones it supports. Systems like IBM Netezza~\cite{singh2011introduction}, PolarDB~\cite{cao2020polardb}, and Ibex~\cite{woods2014ibex} depend on sophisticated and costly hardware like FPGAs and Smart SSDs to perform table scanning inside the storage layer. These systems employ hardware-software co-design techniques to serve their specific use cases. Most of these systems generally follow a clean-slate approach and are built from the ground up specifically tailored for query offloading.

In our approach, we take a programmable storage system, Ceph, and extend its filesystem and object storage layers to allow offloading queries leveraging the extension mechanisms it provides. Storage systems like OpenStack Swift~\cite{swift} and DAOS~\cite{liang2020daos} also provide extension mechanisms via Storelets~\cite{storelets} and DAOS middleware~\cite{middlewaredaos} respectively. We embed Apache Arrow libraries inside the storage layer to build the data access logic. Our approach signifies that storage systems should provide extensive plugin mechanisms so that they can be easily extended to support ad-hoc functionality and do not need to modify legacy code or require a complete rebuild.

%% file: tex/6_conclusion.tex
\section{Conclusion}
In this paper, we present a new design paradigm that allows extending the POSIX interface and the object storage layer in programmable object-storage systems with plugins to allow offloading compute-heavy data processing tasks to the storage layer. We also discuss how different data access libraries and processing frameworks can be embedded in the plugins to build a universal data processing engine that supports different file formats. We present an instantiation of this design implemented using Ceph for the storage system and Apache Arrow for the data processing layer. Currently, our implementation supports reading Parquet files only, but support for other file formats can be easily added since we use Arrow as our data access library. We also discuss two alternative file layout designs for storing Parquet files in Ceph in a fashion similar to HDFS that allow efficient querying. We expose our implementation via a RadosParquetFileFormat API which is an extension of the Arrow FileFormat API. We also discuss some performance evaluations of our implementation and demonstrate that offloading compute-heavy query execution to the storage layer helps improve query performance by making queries faster and more scalable.